# Compact SPICE model for TeraFET resonant detectors

Xueqing Liu, Yuhui Zhang, Trond Ytterdal, and Michael Shur

*Abstract*—This paper presents an improved compact model for TeraFETs employing a nonlinear transmission line approach to describe the non-uniform carrier density oscillations and electron inertia effects in the TeraFET channels. By calculating the equivalent components for each segment of the channel—conductance, capacitance, and inductance—based on the voltages at the segment's nodes, our model accommodates non-uniform variations along the channel. We validate the efficacy of this approach by comparing terahertz (THz) response simulations with experimental data and MOSA1, EKV TeraFET SPICE models, analytical theories, and Multiphysics simulations.

*Index Terms*— TeraFET, terahertz detection, compact model, SPICE

## I. INTRODUCTION

THE terahertz plasmonic field effect transistors (or TeraFETs) have found versatile applications such as THz detectors [1,2], imagers [3,4], and transceivers [5,6]. The effective compact models for TeraFETs have been used for designing and simulating THz components, circuits, and systems [7-10]. Previously developed compact models for TeraFETs have demonstrated qualitative agreement with analytical theories and measured data [11-14]. In this work, we propose an upgraded compact model for TeraFETs achieving improved quantitative alignment with analytical theories and experimental data. The model uses an improved segmentation approach for the device channel [12-14] to accommodate the non-uniform carrier density oscillations and electron inertia effects. For each segment, the model computes the equivalent components for the channel, including conductance, capacitance, and inductance, based on the voltages at the segment's nodes. Consequently, these components vary non-uniformly along the channel. We employ our new approach to improve the previous SPICE model frameworks (MOSA1 [15] and EKV [16]) and validate its effectiveness through comparisons with measured data and previous SPICE modeling results, analytical THz response theories, finite-element simulations using COMSOL Multiphysics software. We observe good quantitative agreement between our new SPICE models, experimental data, analytical theories, and COMSOL simulations, even at resonant detection conditions. The improved TeraFET SPICE model can be effectively used for the design and simulation of THz electronic and optoelectronic devices and circuits.

## II. TERAFET SPICE MODEL

In the presence of THz radiation, field-effect transistors (FETs) with asymmetric boundary conditions demonstrate a notable DC voltage response which results from the rectification of decayed or resonant plasma waves within the FET channel. The hydrodynamic equations governing carrier density in the channel [17,18] analytically describe this THz response. These equations with asymmetric boundary conditions can also be effectively solved using the finite element method. Also, through simulations conducted in COMSOL Multiphysics can describe the THz response and examine plasma wave profiles, thereby gaining valuable insights into the behavior of TeraFETs within the THz frequency domain [19-22].

Previous studies have reported on the compact SPICE models for TeraFET detectors using the model of distributed transmission lines within the channel [13,14]. Fig. 1 shows the equivalent circuit of the multi-segment compact model for TeraFETs, including parasitic capacitances and series resistances. The nonlinear transmission line for the channel is implemented by dividing the channel resistances, capacitances, and inductances for the intrinsic FET into multiple segments. The channel length of each FET segment is $L/N$, where $L$ represents the total channel length and $N$ denotes the number of segments. The Drude inductance, characterizing the electron inertia effect in the channel, is expressed as $L_{drude}=\tau/g_{ch}$, where $g_{ch}=\partial I_d/\partial V_d$ is the channel conductance, $\tau=m\mu/q$ is the electron momentum relaxation time, $m$ is the electron effective mass, and $\mu$ is the electron mobility [23]. In prior models, $g_{ch}$ is calculated based on the voltage at the drain node $d$ and the source node $s$, subsequently divided by $N$ to enable the same $L_{drude}$ for each segment [13,14]. In the enhanced model presented in this paper, the channel conductance for segment $i$ (as shown in Fig. 1) is determined based on the voltages at the respective drain node $d_i$ and source node $s_i$. Consequently, $L_{drude}$ is no longer uniformly distributed along the channel. This varying $L_{drude}$ accounts for the varying channel resistances across segments. Fig. 2 shows a profile for the variation of $L_{drude}$ defined as $\Delta L_{drude} = L_{drude} - L_{drude0}$, where $L_{drude0}$ represents the Drude inductance calculated without THz radiation and $L_{drude}$ is the Drude inductance calculated under THz radiation. Here $\Delta L_{drude}$

(Corresponding author: Xueqing Liu).
Xueqing Liu, Yuhui Zhang, and Michael Shur are with the Department of Electrical, Computer and Systems Engineering, Rensselaer Polytechnic Institute, 110 8th Street, Troy, NY 12180, USA (e-mail: xueqing1988@gmail.com, ericzrpi@gmail.com, shurm@rpi.edu).
Trond Ytterdal is with the Department of Electronic Systems, Norwegian University of Science and Technology, 7491 Trondheim, Norway (e-mail: trond.ytterdal@ntnu.no).

is plotted as a function of the THz radiation frequency and the drain node for each of the 50 segments, which represents different positions along the 90nm TeraFET channel. As seen, $L_{drude}$ is highly nonuniform along the device channel and exhibits resonant features at relatively high frequencies.

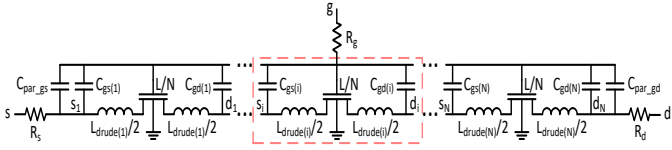

**Fig. 1.** Equivalent circuit of TeraFET compact model. In contrast to previous models, the Drude inductance is a function of the node numbers.

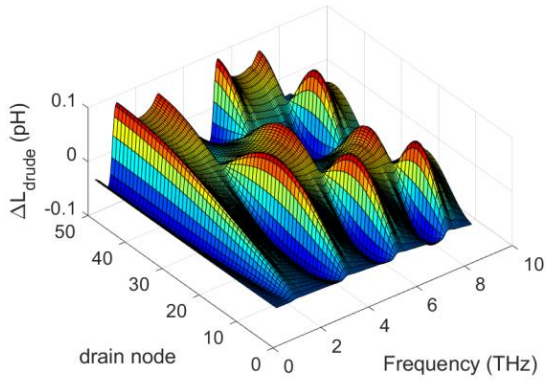

**Fig. 2.** Profile of $L_{drude}$ variation for a TeraFET with 90nm channel length using EKV 50-segment model with varying $L_{drude}$.

To obtain a validated compact SPICE model for TeraFETs, we first adjusted the SPICE parameter values to fit the current-voltage characteristics of an FDSOI MOSFET with a 90 nm channel length simulated in Sentaurus TCAD. Fig. 3 shows the good agreement of I-V characteristics between the MOSFET TCAD model and both the single-channel and multi-segment MOSA1 SPICE models.

The validated MOSA1 SPICE model can be used to simulate the THz response of TeraFETs under open drain boundary condition. We also add the EKV model as the core for the TeraFET SPICE model to compare with experimental data [24]. Fig. 4 shows the comparison of simulated THz response using the single-channel or multi-segment SPICE models and the measured THz response for NMOSFETs with different channel lengths [24]. The good agreement further justifies the use of the SPICE models for TeraFET simulations.

To evaluate the efficacy of the varying $L_{drude}$ approach, we use both MOSA1 and EKV multi-segment SPICE models to compare simulation results with the analytical THz response $\Delta U = V_a^2 f(\omega)/(4V_{gt})$ [17] and the simulated response with COMSOL Multiphysics.

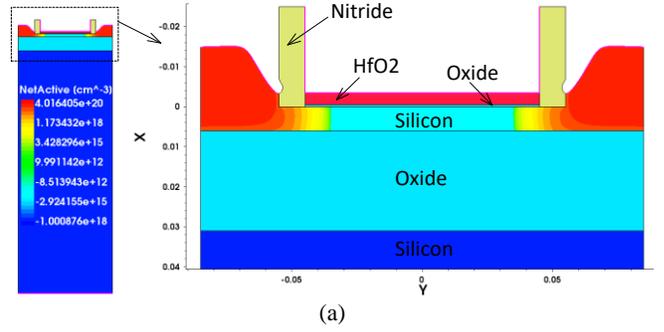

(a)

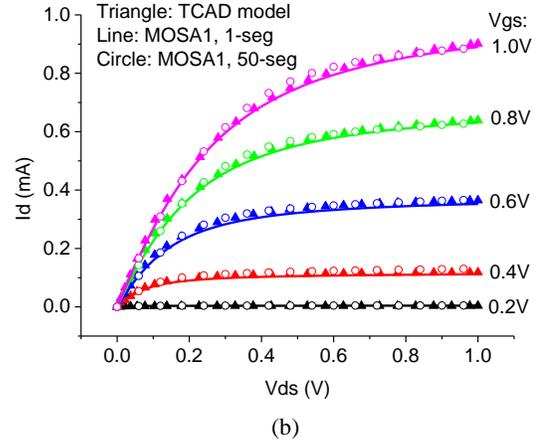

(b)

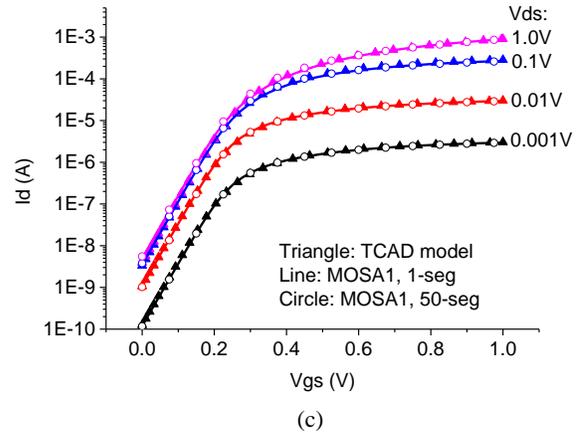

(c)

**Fig. 3.** Schematic of a 90nm FDSOI MOSFET TCAD model (a) and comparison of IV characteristics between the TCAD model and both the single-channel and multi-segment MOSA1 SPICE models: (b) output characteristics and (c) transfer characteristics.

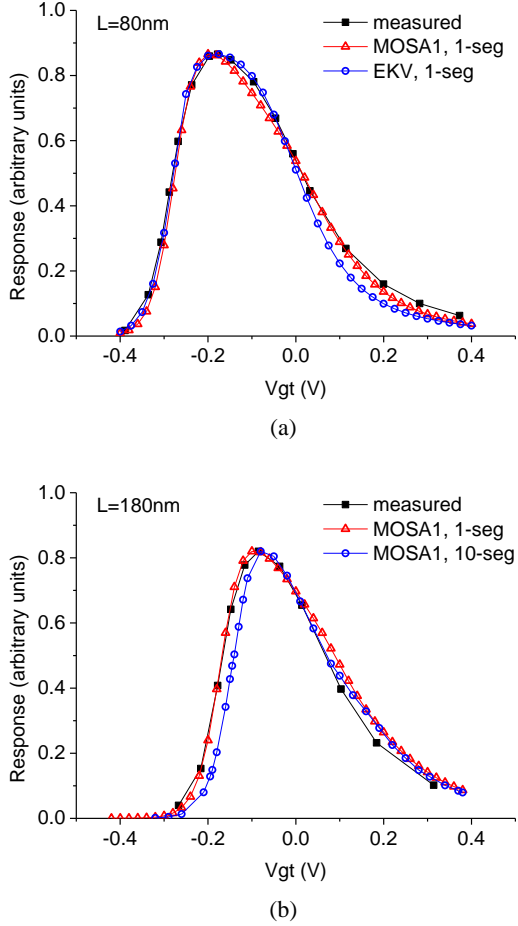

**Fig. 4.** Comparison of simulated THz response using the TeraFET SPICE models and measured THz response for NMOSFETs [24] with (a) 80 nm channel length and (b) 180 nm channel length.

The description of our COMSOL can be found in [19-22]. Fig. 5 shows the THz response as a function of the frequency of the THz signal applied between the gate and source for the TeraFET biased above threshold (gate voltage swing $V_{gt}$ = 0.15V). The comparison employs different approaches which include the analytical THz response equations, COMSOL simulations, and SPICE simulations with multi-segment MOSA1 or EKV models using uniform or varying $L_{drude}$, under different mobility and channel length conditions. The comparison reveals that for conditions involving relatively long channels with high mobility or short channels with low mobility, the improved SPICE models using varying $L_{drude}$ exhibit excellent quantitative agreement with the analytical and COMSOL results. In contrast, the previous SPICE models using uniform $L_{drude}$ only have qualitative agreement with the analytical and COMSOL results. However, for high mobility TeraFETs with shorter channels, the SPICE models using the varying $L_{drude}$ demonstrate higher response than the analytical and COMSOL results but still show much better quantitative agreement than the SPICE models using uniform $L_{drude}$. The overshot THz response in this parametric range might result from the intensified resonant feature of $L_{drude}$, which overestimates the nonuniformity of the device channel.

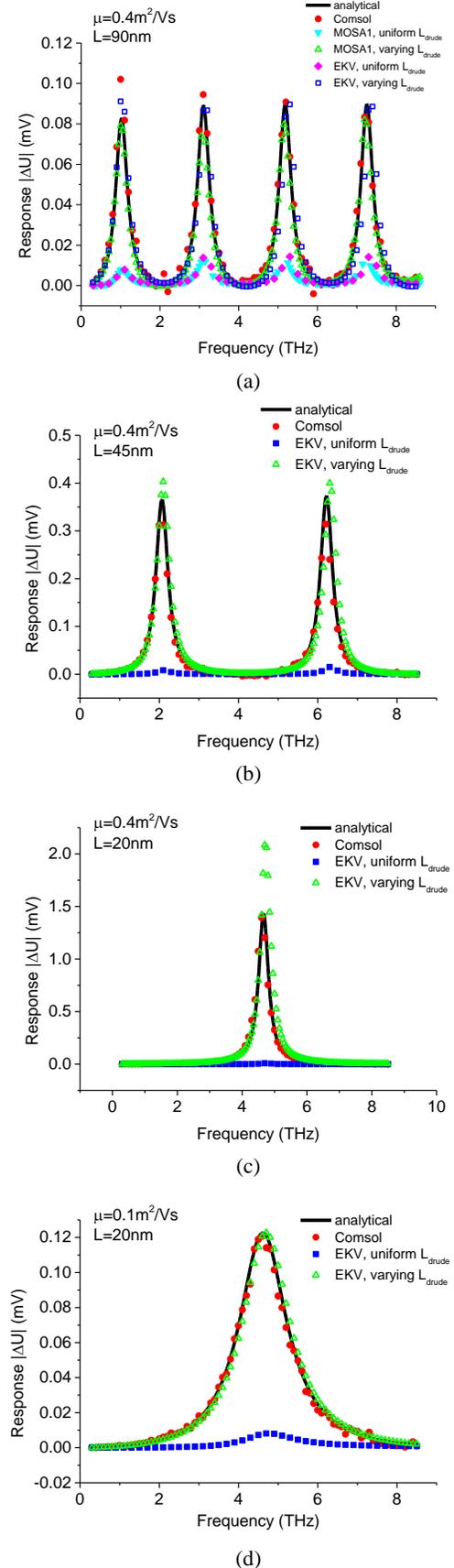

**Fig. 5.** Comparison of THz response between the analytical theory, COMSOL simulation, 50-segment SPICE models with uniform $L_{drude}$ or varying $L_{drude}$ for TeraFETs with (a) 90 nm channel length and 0.4 m$^2$/Vs mobility, (b) 45 nm channel length and 0.4 m$^2$/Vs mobility, (c) 20 nm channel length and 0.4 m$^2$/Vs mobility, and (d) 20 nm channel length and 0.1 m$^2$/Vs mobility.

distributions but different electron density profiles between COMSOL and SPICE simulations. This suggests the dominant role of the drift velocity in determining the THz response under the resonant conditions, as the small-signal condition for electron density is retained.

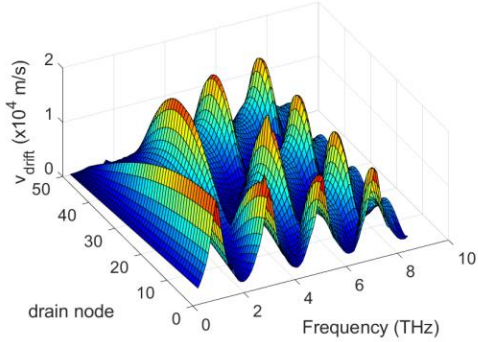

(a)

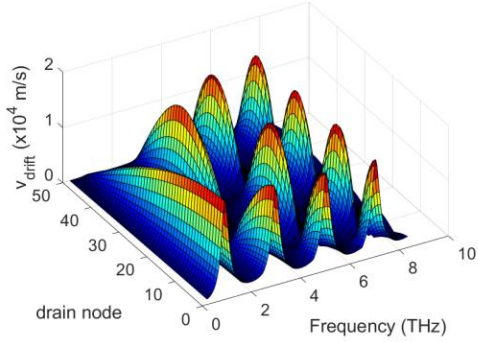

(b)

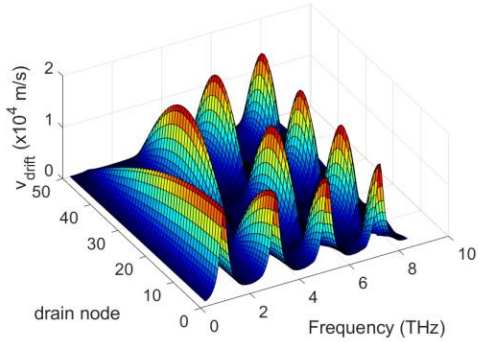

(c)

**Fig. 6.** Profile of drift velocity for 90nm MOSFET ($\mu$=0.4 m$^2$/Vs) under THz radiation: (a) COMSOL simulation, (b) MOSA1 50-segment model with varying $L_{drude}$, (c) EKV 50-segment model with varying $L_{drude}$.

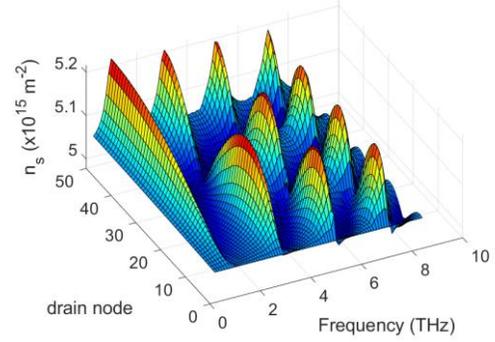

(a)

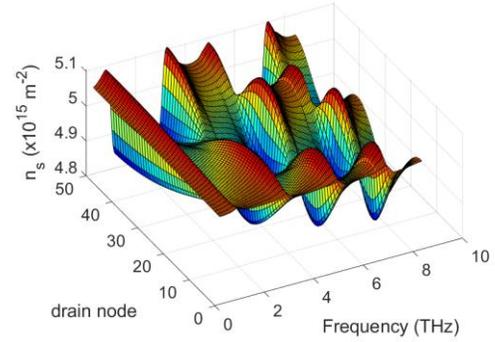

(b)

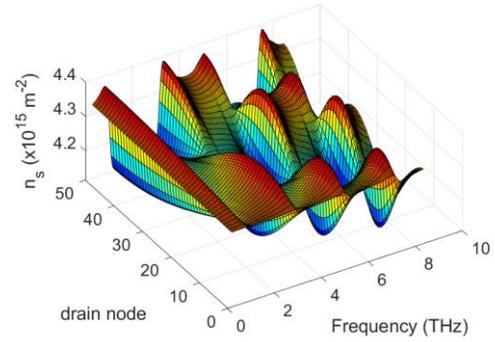

(c)

**Fig. 7.** Profile of electron density in the channel for 90nm MOSFET ($\mu$=0.4 m$^2$/Vs) under THz radiation: (a) COMSOL simulation, (b) MOSA1 50-segment model with varying $L_{drude}$, (c) EKV 50-segment model with varying $L_{drude}$.

Since the THz response stems from the rectification of the nonlinear THz current in the channel and the THz current is induced by the modulation of the carrier drift velocity and the carrier density under the THz radiation [17], it is crucial to investigate the distributions of the carrier drift velocity and the carrier density. Fig. 6 and 7 show the profiles of drift velocity and electron density in the channel for a relatively long channel and high mobility TeraFET, simulated by COMSOL and SPICE using multi-segment MOSA1 or EKV models with varying $L_{drude}$. The comparison shows similar drift velocity

Discrepancies in the electron density profiles may arise from different bias conditions for each segment in COMSOL and SPICE simulations. The analytical THz response expressions and the COMSOL simulations are based on the gradual channel approximation and one-dimensional modeling of boundary conditions, while the SPICE simulations adopt a more complex two-dimensional modeling approach without simplified assumptions.

In the high-mobility or the short-channel low-mobility conditions, TeraFETs exhibit a clear resonant response, where the resonant peaks in the drift velocity profiles may overshadow any difference in the electron density profiles between COMSOL and SPICE simulations when determining the THz response.

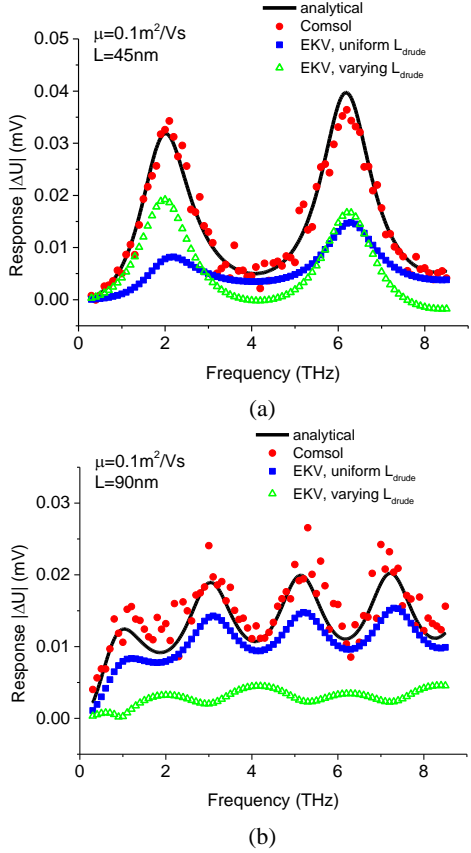

**Fig. 8.** Comparison of THz response between the analytical theory (lines), COMSOL simulation (circles), and 50-segment SPICE model with constant $L_{drude}$ (rectangles) and variable $L_{drude}$ (triangles) for low mobility MOSFET with (a) 45 nm channel length and (b) 90 nm channel length.

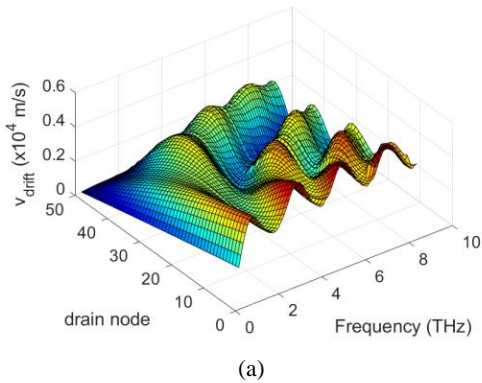

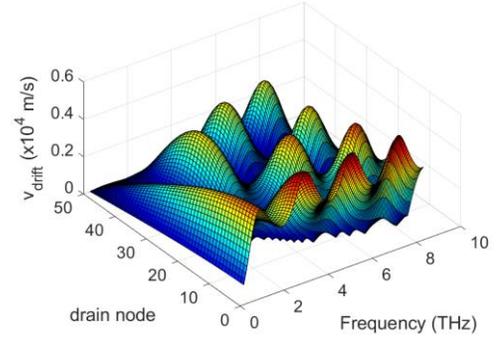

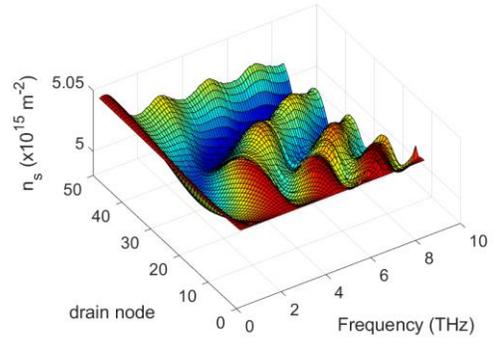

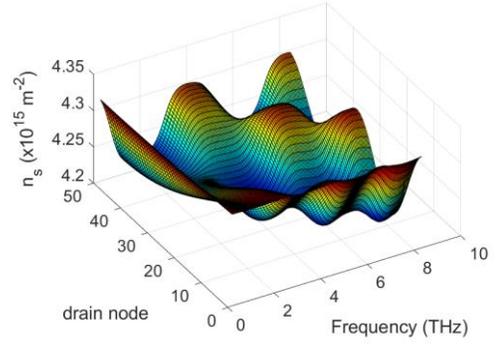

**Fig. 9.** Profile of drift velocity for 90 nm MOSFET ($\mu = 0.1$ m$^2$/Vs) under THz radiation: (a) COMSOL simulation, (b) EKV 50-segment model with varying $L_{drude}$; and profile of electron density in the channel for 90 nm MOSFET ($\mu = 0.1$ m$^2$/Vs) under THz radiation: (c) COMSOL simulation, (d) EKV 50-segment model with varying $L_{drude}$.

Fig. 8 shows a comparison of the TeraFET response among the analytical results, COMSOL simulations, and SPICE simulations using multi-segment EKV models with uniform or varying $L_{drude}$, for TeraFETs with relatively low mobility (0.1 m$^2$/Vs) and channel lengths of 45 nm and 90 nm. For the 45 nm channel length, the SPICE model with varying $L_{drude}$ exhibits a smaller response than the analytical results and COMSOL simulations. However, when the channel length increases to 90 nm, the results of the SPICE model with varying $L_{drude}$ deviate from the analytical results of COMSOL simulations. Fig. 9 shows the profiles of drift velocity and electron density in the channel for the 90 nm low-mobility TeraFET simulated by COMSOL and multi-segment EKV model with varying $L_{drude}$. The comparison also shows similar drift velocity distributions but different electron density profiles between COMSOL and SPICE simulations.

However, for the condition of 90 nm channel length and 0.1 m$^2$/Vs mobility, drift velocity peaks are considerably smaller compared to high-mobility conditions, where the resonant response is evident. Consequently, differences in electron density profiles between COMSOL and SPICE simulations become significant, potentially playing a larger role in determining the THz response.

The above observations indicate that the proposed varying-$L_{drude}$ model is valid only in the strong resonant mode. In this mode, the channel length $L \ll L_{cr}$, where $L_{cr} = s\tau(1+(\omega\tau)^{-1})^{0.5}$ is a critical length signifying the traveling distance of plasma waves, and $s$ and $\tau$ are plasma wave velocity and momentum relaxation time of carriers, respectively [25]. When $L$ approaches $L_{cr}$, both the uniform and varying $L_{drude}$ approaches fail to keep track of the subtle changes in electron inertia, thereby leading to differences observed between the SPICE models and the analytical results or COMSOL simulations in the non-resonant response conditions.

III. CONCLUSION

The enhanced compact model for TeraFETs significantly improves quantitative alignment with measured data, analytical THz response theories, and FEM numerical simulations. By incorporating non-uniform carrier density oscillations and electron inertia effects in the nonlinear transmission line of the channel, our model accurately captures the behavior of TeraFET detectors, particularly in scenarios involving resonant detection. This advanced TeraFET SPICE model can be used for more accurate predictions and optimizations of TeraFET response for THz technology applications.